\documentstyle[12pt]{article}
\newcommand{\s}{$\bar{s}s$}
\newcommand{\p}{$\bar{p}p$}
\newcommand{\f}{$\phi$}
\newcommand{\an}{annihilation}

\begin{document}
\setcounter{page}{0}
\begin{titlepage}
\begin{flushright}
{CERN--TH.7326/94 \\
TAUP--2177/94}\\
hep-ph/9412334
\end{flushright}
\medskip
\begin{center}
{\Large\bf Abundant $\phi$-meson production in $\bar{p}p$ annihilation
\\ at rest and strangeness in the nucleon  \\~\\}

\end{center}
\medskip
\begin{center}
{\large J. Ellis$^{a}$, M. Karliner$^{b}$,
D.E. Kharzeev$^{a,c,}$\footnote{On leave of absence from Moscow State
University} and
M.G. Sapozhnikov$^{d}$}
\end{center}
\vskip1cm
\begin{center}
$^{a)}$ Theory Division, CERN, Geneva, Switzerland\\
\vskip0.3cm
$^{b)}$ School of Physics and Astronomy,
Tel--Aviv University, Tel--Aviv, Israel\\
\vskip0.3cm
$^{c)}$ Physics Department, University of Bielefeld, Bielefeld, Germany\\
\vskip0.3cm
$^{d)}$ Joint Institute for Nuclear Research, Dubna, Russia
\end{center}
\vskip2.5cm
\begin{abstract}
A large apparent violation of the OZI rule has recently been found in
 many channels in $\bar{p}p$ \an\ at LEAR.
An interpretation of these
data in terms of the ``shake-out" and ``rearrangement" of an intrinsic \=ss
component of the nucleon wave function is proposed.
This gives  a channel-dependent, non-universal modification of the
na\"{\i}ve OZI prediction.
Within this approach, we interpret the strong excess of $\phi$ production in
S-wave \=pp annihilations in terms of the polarization of the nucleon's
 \=ss component indicated by
deep inelastic lepton-nucleon scattering experiments.
 This interpretation
could be tested by measurements of the $f_2'(1525)/f_2(1270)$ production ratio
in P-wave annihilations and by experiments with polarized beams and
polarized targets.
We also propose a test of the
intrinsic strangeness hypothesis in $\phi$ production in high-momentum
transfer processes,
via a difference in constituent counting rules from gluon-mediated
production.
\end{abstract}
\vskip1.8cm
CERN--TH.7326/94\\
TAUP--2177/94\\
December 1994
\end{titlepage}
\newpage
\section{Introduction}

According to the na\"{\i}ve constituent quark model, the
proton wave function contains just
two u-quarks and one d-quark. This model gives a good general picture of
hadron structure at large distances, but probes of shorter distances
and larger momentum transfers
reveal more constituents,
including a sea of \=qq (q=u,d,s) pairs and gluons.
These evolve with momentum,  in agreement with
perturbative QCD, but there are indications that the nucleon wave function
contains some
\=ss pairs
already at large distances or small momentum transfers, in the
non-perturbative regime
(for reviews, see
\cite{Ell.89}-\cite{Dec.92}). This should not be unexpected, since
non-perturbative QCD effects cause \=ss pairs to be present in the vacuum:
$< 0\left|\bar{s}s \right|0> \neq 0$, and soluble two-dimensional models
\cite{2dim} also
predict the existence of non-perturbative \=ss pairs in the nucleon
wave function. This possibility is also consistent with analytical \cite{GL.91}
and lattice \cite{lat} determinations
of the $\pi-N$ $\sigma-$ term.
\vspace {0.3cm}

Interesting experimental evidence for such an \=ss component
comes from the apparent violation of the
Okubo-Zweig-Iizuka (OZI) rule in the production of $\phi$ mesons in
 different reactions.
According to the OZI rule \cite{OZI}, diagrams with disconnected quark lines
 should
be negligible. Production of the $\phi$ meson is a particularly
sensitive probe of the OZI rule because the
$\phi$ is almost a pure \=ss state, containing just a small admixture
of \=uu+\=dd associated with a small deviation
$\delta = \Theta - \Theta_i $  from the
    ideal mixing angle $\Theta_i=35.3^0$.
The OZI rule was used
\cite{Lip.76}
to predict
\begin{equation}
R = \frac{\sigma(A+B \rightarrow\phi X)}{\sigma(A+B\rightarrow\omega X)}
= \tan^2{\delta}\cdot f \label{R}
\end{equation}
for any initial-state hadrons A,B and final-state hadrons not containing
strange quarks, where $f$ is a kinematical phase space factor.
Using the theoretically-favoured
quadratic Gell-Mann-Okubo
mass formula, one finds  $\Theta=39^0$
and hence
$R = 4.2\cdot10^{-3}f$.
As we see in Table 1, many past experiments found an apparent excess of R
above this OZI prediction, though this was not very dramatic:
$R\leq(10-20)\cdot10^{-3}$.
\vspace {0.3cm}

The significance of such an excess can
be inferred from the parameter
\begin{equation}
Z = \frac{ M(A+B\rightarrow \bar s s + X)}
{[M(A+B\rightarrow \bar u u +X) + M(A+B\rightarrow \bar d d +X)]/\sqrt{2}}
\label{z}
\end{equation}
which measures OZI-breaking in the amplitudes
$M(A+B\rightarrow \bar{q}q +X)$, leading to the estimate
\begin{equation}
R = \left( \frac{ Z+ \tan \delta}{1-Z\tan \delta} \right)^2 \cdot f \label{z1}
\end{equation}
Values of
$\left| Z\right|$ are also shown in Table 1, where we see that previous
experiments indicate
$\left| Z\right|\leq 0.1$ .
    However, as we discuss later in this paper, much larger apparent
violations of the OZI rule,
$\left| Z\right|\leq 0.2-0.4$,
have recently been found in \=pp annihilation
at the Low Energy Antiproton Ring (LEAR) at CERN, as reviewed in Table 2.
\vskip0.3cm

The main purpose of this paper is to pursue the interpretation of these
data in terms of the ``shake-out" and ``rearrangement" of an intrinsic \=ss
component of the nucleon wave function as illustrated in Figs.1a,b,
respectively.
These mechanisms provide {\it a channel-dependent, non-universal
modification} of the
na\"{\i}ve OZI prediction (\ref{R}), and we propose tests of our
interpretation.
This should not be considered a violation of the OZI rule, because it
does not involve disconnected quark diagrams,
but rather connected diagrams involving higher Fock--space components in the
nucleon wave function. Within this approach, we interpret the strong excess of
 $\phi$ production in
S-wave \=pp annihilations in terms of the negative polarization of the
nucleon's
 \=ss component indicated by EMC and subsequent results \cite{EMC}.
 This interpretation
could be tested by measurements of the $f_2'(1525)/f_2(1270)$ production ratio
in P-wave annihilations, where there may be an enhancement over the
na\"{\i}ve OZI expectations, and by experiments with polarized beams and
polarized targets.
We also mention a possible test of $\phi$ production within the
intrinsic strangeness hypothesis in high-momentum transfer processes,
via a difference in constituent counting rules from gluon-mediated
production.

\section{The ``Shake--out" and ``Rearrangement" of
Intrinsic Strangeness in the Nucleon}

    We start with some general considerations on $\phi$ production
via an intrinsic strangeness component of the nucleon.
    We adopt the following notation for the decomposition of the
proton wave function:
\begin{equation}
|p> = x \sum_{X=0}^{\infty}|uudX>  +
    z \sum_{X=0}^{\infty}|uud\bar{s}sX> , \label{wf}
\end{equation}
where $X$ stands for any number of gluons and light \=qq pairs,
and the condition $|x|^2 + |z|^2 = 1$ holds if we neglect
the admixture of more than one $\bar s s$ pair.
We consider two processes which are allowed by the OZI rule if such
an intrinsic strangeness component is present, namely the ``shake--out"
illustrated in Fig.1a and the ``rearrangement" illustrated in Fig.1b.
A ``shake--out" amplitude for $\bar{p}p$ annihilation into a state
with either hidden strangeness such as the $\phi$ or open strangeness, is
 given generically by
\begin{equation}
M(\bar{p} p \rightarrow \bar s s + X) \simeq 2 Re(x z^*)\ P(\bar s s),
\label{sh}
\end{equation}
where $P(\bar s s)$ is a projection factor which depends in
particular on the final state considered: $\phi$, $f'_2(1525)$, non--resonant
$\bar{K}K$ pair, etc., though
there may also be some dependence on the initial $\bar{p}p$ state
considered. A ``rearrangement" amplitude for producing a particular
\s\ state
is given generically by
\begin{equation}
M(\bar{p} p \rightarrow \bar s s + X) \simeq |z|^2\ T(\bar s s), \label{re}
\end{equation}
where the factor $T(\bar s s)$ will in general depend quite strongly on both
 the initial
and final spin states, since the $\bar{s}$ and $s$ come from different
initial--state particles. For example, this mechanism may be expected
to be unlikely to give
a P--wave strangeonium state such as the $f'_2(1525)$ if the $\bar{p}p$
annihilation takes place from an S--wave state.
The corresponding amplitudes for
 producing a light
\=qq state with the same quantum numbers
are not directly related to (\ref{sh}) and (\ref{re}), but would be given
generically by
\begin{equation}
M(\bar{p} p \rightarrow \bar q q + X) = |x|^2\ \left[P(\bar q q)\ or\
T(\bar q q)\right],
\end{equation}
in analogous notation.
\vskip0.3cm

Defining Z as in equation (\ref{z}), and assuming
that the factors $P$, $T$ are
similar  for light and strange quarks, we find
\begin{equation}
|Z| = 2 \left|{z \over x}\right|
= 2 {|z| \over \sqrt{1-|z|^2}} \label{sha}
\end{equation}
for the ``shake--out" diagram and
\begin{equation}
|Z| =  {|z|^2 \over |x|^2} = {|z|^2 \over {1 - |z|^2}}   \label{rea}
\end{equation}
for the ``rearrangement" one. In particular, the linear dependence (\ref{sha})
 for shake-out processes allows $|Z|$ to be relatively large, even if
$|z|^2$ is small.
If we choose even  one of the most striking values in Table 2, namely
$|Z(\phi\pi/\omega\pi)| = 0.24\pm0.02$, we see from (\ref{sha})-(\ref{rea})
that the \=ss admixture
needed  in the proton wave function
is in the range
\begin{equation}
0.01 \leq |z|^2 \leq 0.19   \label{per}
\end{equation}
         Such an admixture is not incompatible with data on open
strangeness production in \=pp annihilation at rest, as we now
discuss.
\vspace {0.3cm}

    Our hypothesis that there are \=ss pairs ``stored" in nucleons,
that can be ``shaken out" or ``rearranged" in \=pp annihilation,
can be confronted with
the available data on strange particle production in annihilation
at rest.
Summing all measured channels containing kaons and correcting
for unseen modes, the kaon yield is found to be  \cite{Bat.90}
\begin{equation}
Y_K=(4.74\pm 0.22)\% \label{yk}
\end{equation}
which is consistent with the independent determination of the yield of
annihilations into pions alone \cite{Bal.66}
\begin{equation}
Y_{\pi}=
(95.4\pm 1.8)\%.
\end{equation}
    Among the contributions to $Y_K$ are the ``shake-out" and
``rearrangement" processes we
discuss, and the creation of new \=ss pairs in the annihilation final state.
Examples of the latter are provided by the kaonic decay modes of
non-strange mesons such as the
$ f_2(1270)$,  $a_2(1320)$ and $b_1(1235)$.
Combining  their production branching ratios with their
probabilities for decay into final states containing $\bar{K}K$,  one
estimates a contribution to $Y_K$ of
$0.4-0.5\%$ of all annihilations.
Most of the remainder of $Y_K$ could be due to
  ``shake-out" and ``rearrangement"
contributions of the order of 4 \%, which is comparable with (\ref{per}).
Indeed, it follows from (\ref{sh}) that
\begin{equation}
Y_K = 4 |x|^2 |z|^2 cos^2\varphi = 4 (1-|z|^2) |z|^2 cos^2\varphi,
\label{yield}
\end{equation}
where $\varphi$ is the relative phase of the $x$ and $z$ (complex)
coefficients. The value $Y_K\simeq 4\%$ therefore leads to the limit
\begin{equation}
|z|^2 \geq 0.01, \label{limit}
\end{equation}
which is consistent with (\ref{per}).
The relative magnitude
of the contribution from the creation of new $\bar{s}s$ pairs could in
principle
 be probed by comparing
$Y_K$  (\ref{yk}) with the kaon yield in $\pi\pi$ scattering at the
same centre-of-mass energy.
The $\bar{K}K$ yield due to ``shake-out" or ``rearrangement" in \=pp
annihilation
could also have phase-space distribution different from that of
\=ss pair creation. The latter would tend to be central and with low relative
momenta, whereas ``shake-out" contributions could carry large momenta and
emerge in the forward and backward directions from
\=pp annihilation in flight.
Indeed, the observed \cite{Tan.85} backward peak in
$\bar{p}p\rightarrow K^-K^+$ at $p=0.5~GeV/c$ was interpreted \cite{Ell.89}
as evidence for strange quarks in the proton.
Recent experiments at LEAR and KEK confirm existence of the strong
backward peak at \=p momenta $p< 1~ GeV/c$
\cite{Has.92},\cite{Tan.90}.
\vskip0.3cm

Additional evidence for the intrinsic strangeness picture comes from the
experimental data on double $\phi\phi$ production taken by
 the JETSET Collaboration at LEAR. Na\"{\i}vely, one would expect
\cite{Dov.92} a cross section of the order of
\begin{equation}
\sigma(\bar{p}p\to \phi\phi) = tan^4\delta\  \sigma(\bar{p}p\to \omega\omega)
\sim 10\ nb, \label{jet}
\end{equation}
if both $\phi$'s were produced by independent OZI--violating couplings.
The cross section measured \cite{Mac.93}
at 16 different
antiproton momentum settings spanning the region between 1 and 2 GeV/c
is however about 1500 nb, i.e. it exceeds the na\"{\i}ve OZI rule
estimate given above by at least two orders of magnitude.
The intrinsic strangeness model easily accommodates this
experimental observation, which could be due to either
``shake--out" or ``rearrangement", with an amplitude $\sim |z|^2$.
The limit (\ref{limit}) would imply then
\begin{equation}
\sigma(\bar{p}p\to \phi\phi) = {|z|^4 \over {|x|^4}}\
\sigma(\bar{p}p\to \omega\omega) \geq 250\ nb.
\end{equation}
The experimental value of the cross section corresponds to
$|z|^2 \simeq 0.025$.
  The connected diagrams involving intrinsic strange quarks
  will likely mask any possible glueball resonance
contributions which are expected to be dominant among the disconnected
diagrams.
No such resonances are observed at present \cite{Mac.93}.

\section{Data on $\phi$ Production in \p\ Annihilation at Rest}

    After this preparatory discussion of the intrinsic \=ss component
of the nucleon and the ``shakeout" and ``rearrangement" processes, we now
discuss specific features
of $\phi$ production in \=pp annihilation at rest. The fact that the ratio
of $\phi$ and $\omega$ yields in various individual channels exceeds the
prediction of the na\"{\i}ve OZI rule has been known for some time. In
 particular,
the bubble chamber data of
\cite{Biz.74}-\cite{Bet.67} on $\bar{p}n$ annihilation could be combined to
 estimate
$ R_{\pi^-}= B.R.(\phi\pi^-)/B.R.(\omega\pi^-) = (83\pm25)\cdot10^{-3}$.
However, the
statistics was limited to  54 events in the $\phi\pi^-$ channel \cite{Biz.74}.
    We also mention the value of
\begin{equation}
R={g^2_{p\bar{p}\phi} \over {g^2_{p\bar{p}\omega}}} = 0.211 \div 0.276.
\end{equation}
inferred from analyses of the proton vector isoscalar form factor
 \cite{Hoh76},
\cite{Dub} in which the $\bar{N}N\phi$ and $\bar{N}N\omega$ coupling constants
were treated as free parameters (see also \cite{Jaf89}).
This value is compared with measurements of $\phi/\omega$ production ratios
in Fig.2 \footnote{ We note here in passing that a triply-strange baryon
resonance has been seen \cite{Ast.88} in the channel
 $\Omega^*\rightarrow\Omega\pi\pi$,
which is na\"{\i}vely OZI-disallowed, but would be allowed if there is
a higher $|sss\bar{q}q>$ Fock-state component in the $\Omega$ wave function.}.
\vspace {0.3cm}

As we now review, the advent of high-statistics experiments at LEAR has
brought a new
era in \=pp annihilation studies, with larger excesses in $\phi$ production
above the na\"{\i}ve OZI prediction established by the ASTERIX \cite{Rei.91},
Crystal Barrel \cite{Fae.93} and OBELIX \cite{Abl.93} collaborations,
as shown in Fig.2
and Table 2.

\vspace {0.3cm}
       The ASTERIX collaboration \cite{Rei.91}
has measured the ratios
of $\phi X$
and $\omega X$ final states with $X= \pi ,\eta, \omega,\rho, \pi\pi$
in
S- and P-wave \=pp annihilation. As seen in Table 2, the experimental
values of R in different S-wave annihilation channels are mostly higher
than the OZI prediction by a factor of 2 to 8.
The most striking variation in this trend is the very large enhancement in
the case
$X=\pi$  in
S-wave annihilations, by a factor 20 or so.
However, in P-wave annihilations there is no corresponding enhancement
in the cases $X=\pi$ and $X=\eta$, and only modest (if any)
enhancement in the $X=\rho,\pi\pi$ channels, as can be seen by comparing the
yields from the different types of target.
\vspace {0.3cm}

    The Crystal Barrel collaboration
\cite{Fae.93}
has measured the ratios of $\phi X$ and $\omega X$ for $ X  = \pi^0   ,\eta$,
$\pi^0\pi^0$ and $\gamma$ final states in
\=pp annihilation in liquid hydrogen. They confirmed
the ASTERIX observation of a  large deviation
from the na\"{\i}ve OZI rule (\ref{R}) in the case $X=\pi^0$, whilst the
$X=\eta,\pi^0\pi^0$ cases deviate only slightly (if at all)
from (\ref{R}).
However, their most striking result was an extremely large ratio in the
case $X=\gamma$, which they found to be about  100 times
higher
than the na\"{\i}ve OZI prediction.
\vspace {0.3cm}

    The OBELIX collaboration has complemented these results by measuring
the  $\phi X /\omega X$ ratios in \=n  annihilations in liquid
hydrogen \cite{Luc.93}, and in \=p annihilation on gaseous
deuterium at different momenta of the spectator protons \cite{Abl.93}.
Large enhancements over the na\"{\i}ve OZI prediction (\ref{R}) were found in
both the cases $X=\pi^{\pm}$.
\vskip0.3cm

    We draw the reader's attention to the following salient features
of $\phi$ production in nucleon annihilation at rest.

1. The channels
$\phi\pi$,
$\phi\gamma$ exhibit
strong enhancement over the na\"{\i}ve
OZI prediction (\ref{R}), whereas there are only smaller
enhancements in the channels
$\phi\rho$, $\phi\omega$, $\phi\pi\pi$ and no evidence of any
enhancement in $\phi\eta$.

2. The amount of apparent OZI violation in annihilations at rest is
much higher than what is seen in
 $\pi p$ or pp
scattering and higher-energy \=pp annihilation, as shown in  Table 1 and
Fig.2.

3. In cases where the initial \=pp state is known, the large
enhancement of $\phi\pi$
appears to be restricted to the S-wave, with no large
deviation from the na\"{\i}ve
prediction (\ref{R}) in any P-wave annihilation channel.

Any model of the large rate of $\phi$ production in \=pp annihilation at rest
should be able to predict or accommodate these three channel-dependent
features.

\section{Alternative Models of $\phi$ Production}

    It has been suggested \cite{Dov89}
that the enhancement of $\phi$ meson production
in certain $\bar{N}N$ annihilation channels might be due to
resonances.
Specifically, if there existed a
vector ($J^{PC}=1^{--}$) $\phi\pi$ resonance  close to
the $\bar N N$ threshold,
it might be
possible to explain the selective enhancement of the  $\phi\pi$
yield in S-wave  annihilation, and the relative lack of
 $\phi$'s in P-wave annihilation.
    The best candidate for such a state is
one
with mass $M=1480\pm40$ MeV,  width $\Gamma=130\pm60$ MeV and
quantum numbers $I=1,~J^{PC}=1^{--}$, which was observed
\cite{Bit.87} in the $\phi\pi^0$ mass spectrum
in  the reaction $\pi^- p \rightarrow K^+ K^- \pi^0 n$ at 32.5 GeV/c,
and dubbed the C-meson.
\vspace {0.3cm}

However,
this resonance cannot explain the even larger enhancement recently observed
in the $\phi\gamma$ channel, which is a final state with different quantum
numbers. Moreover, the experimental
status of the C-meson is unconfirmed. Although some  experiments
have found  indications for its existence
(for a review, see \cite{Lan.88}), others have not.
In particular, the ASTERIX collaboration \cite{Rei.91} has established an
upper limit of  $3\cdot10^{-5}$ on the production of any $\phi\pi$ resonance
in \=p annihilation in a hydrogen gas target, and the
Crystal Barrel collaboration  has not seen the C-meson among
($\phi\pi^0)\pi^0$ final states
in \=p annihilation in
a  liquid hydrogen target
\cite{Bra.92}.
The predicted \cite{Dov89} isoscalar partner of the C-meson which
should couple to the $\phi\eta$ channel  also was not observed, and no
deviation from the na\"{\i}ve OZI rule has been detected in this mode
(see Table 2).
Therefore we discard the resonance interpretation of the
apparent violation of the OZI rule.
\vspace {0.3cm}

    Alternatively, it has been suggested \cite{Loc.93},\cite{Buz.93}
that the $\phi$ mesons in
the  $\phi\pi$ final state might be due to final-state interactions of the
K and \=K in the $K^*\bar{K} + \bar{K^*}K$  channel that dominates the
$KK\pi$ final state:
$$\bar p
p\rightarrow K^*\bar{K}+\bar{K^*}K\rightarrow\phi\pi$$
One interesting test of this mechanism would be to see whether
isospin I=1 amplitude of
$\bar p
p\rightarrow K^*\bar{K}+\bar{K^*}K$ is larger in S-wave annihilation than
in P-wave annihilations. If not, the large $\phi$ excess in S-wave
annihilations  could not be explained.
However, the concrete calculations of $\phi$ production \cite{Loc.93},
\cite{Buz.93}
 made in different channels
fall below the experimental data by factors of 2 to 6 or more.
Moreover, strong cancellations are expected between different hadronic loop
amplitudes \cite{Lip.91}, \cite{Gei.91}, so these calculations can only be
 considered as upper limits at
best. Furthermore, as  was pointed out in \cite{Kon.93},
the rescattering model cannot explain why the ratio
$\phi\pi\pi/\omega\pi\pi$ is smaller than $\phi\pi/\omega\pi$, despite
the fact that the $K^*\bar{K}^*$ final state is as copious as
$K^*\bar{K} + K\bar{K^*}$ in $p\bar{p}$ annihilation. Therefore
the rescattering interpretation of the apparent violations of the
OZI rule meets serious difficulties.

\section{ Polarization of the Intrinsic Strangeness}

As already mentioned above,
we seek to explain $\phi$ enhancement
in terms of the intrinsic strangeness component in the long--wavelength
nucleon wave function. The key question to be addressed by this explanation
is the channel dependence of the $\phi$ enhancement. As has been pointed out
elsewhere \cite{Ell.89},\cite{Iof.90},\cite{Hen.92}, the $\bar{s}s$ component
of
 the
nucleon is expected to exhibit
momentum and spin correlations that cause couplings to $\bar{s}s$ meson
final states to be channel--dependent in general. We now construct a model
for one aspect of this channel dependence suggested by experimental results
on the polarization of strange quarks and antiquarks in the proton.
Results from the EMC, SMC, E142 and E143 collaborations
\cite{EMC} indicate that they have a net polarization
opposite to the proton spin \cite{Ell.93a}:
\begin{equation}
\Delta s \equiv \int\limits_{0}^{1} dx[ q_{\uparrow}(x) -
q_{\downarrow}(x) + \bar q_{\uparrow}(x)-
\bar q_{\downarrow}(x)] = - 0.08\pm0.03.
\end{equation}
For the sake of simplicity, let us consider the possibility that all the $s$
and
 $\bar{s}$
in the proton are polarized negatively, and none positively.
The simplest wave function of this higher Fock-state component, consistent with
the parity and spin constraints, corresponds to a spin-triplet S-wave
$|\bar{s}s>$ state which is in a P wave relative to the ``normal"
$S={1/2}$ light $|uud>$ component. The projection of the relative
orbital momentum on the
spin axis in this model is $L_z=+1$, so that the total angular momentum
of the nucleon remains ${1/2}\,\,$\footnote{We do
not discuss here dynamical effects
which can be responsible for such a wave function.}.
In this case, ``shake--out" into the $^3S_1$ strangeonium state, namely the
$\phi$, is likely to dominate over other hidden strangeness states such as
the P--wave $f'_2(1525)$.
Now let us consider $\bar{p}p$ annihilation from a spin--triplet initial state,
in which the $\bar{p}$ and $p$ spins are parallel.
In this case, $\bar{s}$ and $s$ quarks in both nucleons
 ``rearranged" during the annihilation will also have parallel spins,
as in the na\"{\i}ve quark model wave function of the $\phi$ meson.
Further, if the $p\bar{p}$ initial state is S--wave, the \s\ pair will
probably also be in an S--wave also as in the \f\ meson. Therefore, we
expect maximum enhancement of \f\ production in the $^3S_1$ channel,
as observed in the $\phi\pi$ final state. This model predicts weaker
enhancements in the $^1S_0$ channel, as observed.
\vskip0.3cm

This model also suggests qualitatively why \f\ production may be enhanced more
in \p\ \an\ at rest than in the other \p , $pp$, and $\pi p$ interactions.
The reason is that higher--energy collisions involve an increasing mixture
of partial waves, implying that the S--wave state ``rearrangement"
that favours \f\ production
becomes progressively more diluted.
A corollary of this observation is
that we would expect the large enhancement of $\bar{N}N \to \phi\pi$ to
diminish as the centre--of--mass energy is increased.
\vskip0.3cm
Thus our heuristic polarization model predicts or accommodates the three
salient features of the data noted earlier. However, it should be emphasized
that our model is idealized. The approximation that all the $\bar{s}$ and $s$
inside a proton are polarized antiparallel to its spin is very crude, the
polarizations could be altered during the ``shake--out" or
``rearrangement"  processes, and diagrams like
those in Fig.1b could make contributions to the production of \f\ 's and other
\s\ mesons that are independent of the initial spin state.
\vspace {0.3cm}

The reader might expect that the $\phi\eta/\omega\eta$ ratio would be
enhanced at least as strongly as the $\phi\pi/\omega\pi$ ratio, since the
favoured $^3S_1$ initial \p\ state also contributes. However, in this channel
there
are additional connected quark diagrams, (shown in Fig.3) because the $\eta$
has a substantial ($\approx50\%$) \s\  -component. Fig.3a features the
rearrangement of two \s\ pairs from the initial nucleon wave functions,
and would require a spin-flip of at least one of the \=s or s to produce
the $\phi\eta$ final state from the $^3S_1$ initial state, according to our
idealized model. Fig.3b involves the annihilation of one \s\ pair and
the subsequent creation of a new \s\ pair, and may be at least as important
as Fig.3a.
\vspace {0.3cm}

    The diagrams in Fig.3 will interfere with the diagrams of Fig.1
that contribute to $\phi\pi$ production in our model. We do not know
\underline{a priori} whether this interference is constructive or
destructive, enhancing or suppressing the yield of $\phi\eta$ relative
to $\omega\eta$. However, the fact the \s\ components of the $\eta$ and
$\eta'$ wave functions have opposite signs means that the interferences in
the $\phi\eta$ and $\phi\eta'$ production amplitudes must have opposite
signs, enhancing one while suppressing the other. The data in Table 2
tell us that the $\phi\eta$ channel is suppressed, enabling us to predict
that the $\phi\eta'$ channel should be enhanced. Since both $\phi$ and
$\eta'$ are heavier than the proton, this prediction can only be checked
in \p\ \an\ in flight. However, it should be noted that the favoured S-wave
channel will be diluted for in-flight annihilation.

\vskip0.3cm

    It would be interesting to explore the
dependence of the amount of the apparent OZI violation not only on
the spin of the initial state but, for instance, on the momentum transfer.
In an early experiment on \f\  production in $\pi^{\pm} N \to \phi N$
 interaction
\cite{Coh.77} it was found that the $d\sigma/dt$ distribution of \f\
production
at large $t$  differs significantly from the one for $\omega$-meson, leading to
the increase of $\phi/\omega$ ratio at large $t$. This effect was especially
marked for unnatural-parity exchange.\footnote{We thank E.Klempt and
C.Strassburger who brought our attention to these results.}
The production of the $\phi$ in antiproton annihilation at rest
 also seems to exhibit a dependence of the apparent OZI rule
violation on the momentum transfer.  In Fig.4 the dependence
of the ratios $R=\phi X/\omega X$ in different reactions
is shown as a function of momentum transfer in $\bar{p}p\to \phi X$ (see
also Fig.5).
The deviation from the na\"{\i}ve OZI rule prediction increases with
momentum transfer.
\vskip0.3cm

However one should be cautious in interpreting the dependence in Fig.4
as a rigorous proof that the apparent OZI rule violation increases with
momentum transfer. In the two-body antiproton annihilations at rest:
$\bar{p}p\to \phi(\omega) + X$, the momentum transfer to the $\omega$
is always higher than the momentum transfer to the $\phi$. It is possible
to compare $\phi$ and $\omega$ production at the same momentum transfer
in annihilation in flight or in the $\phi(\omega)\pi\pi$ channel for
annihilation at rest.
\vskip0.3cm

    The intrinsic strangeness model can also explain rather naturally
 a number of experimental facts on strange baryon
production. For instance, it is well known \cite{Bar.93} that in the reaction
\begin{equation}
\bar{p} + p \to \Lambda + \bar{\Lambda} \label{ll}
\end{equation}
the spins of the $\Lambda$'s exhibit strong correlations. Although both
spin singlet and triplet final states are possible, the spin singlet
fraction is zero within statistical errors. The spin
of the $\Lambda$ is largely carried by  the spin of the strange quark, so the
\s\  pair in the final state of (\ref{ll}) must mostly have parallel spins.
This could
naturally be expected in the intrinsic strangeness model: a spin-triplet
\s\  pair in the
proton or antiproton may
 simply
dissociate into a $\Lambda \bar{\Lambda}$ pair, conserving their spin
correlation, which leads to a spin-triplet final state for the two
hyperons.

\section{Possible Tests of the Model}

Possible tests of our model include checks on the spin-dependences of
amplitudes that violate the na\"{\i}ve OZI rule, and on their momentum
transfer dependences.
\vskip0.3cm

1.\ The arguments for $\phi\pi$ enhancement in
production from the $^3S_1$ initial state can be extended to other
\s\ resonances, in particular to production of the $f_2'(1525)$ compared
to the $f_2(1270)$. Using the quadratic mass formula, as used to obtain
the na\"{\i}ve OZI estimate (\ref{R}), we obtain
\begin{equation}
R' = f'_2(1525)/f_2(1270) = 16\cdot10^{-3} \label{f1}
\end{equation}
before applying phase space corrections. This may actually be an
overestimate: the OZI--forbidden decay $f'_{2}\to\pi\pi$ has been seen
at a low rate relative to the OZI--allowed $\bar{K}K$ decay mode,
corresponding to
\begin{equation}
R' = f'_2(1525)/f_2(1270) = 3\cdot10^{-3}.     \label{f2}
\end{equation}
The $f_2'(1525)$ was not seen by bubble chamber
experiments in \an s at rest, which gave an upper limit on the yield for
$\bar{p}p\to \pi^0 f'_2$ of
$3.8\cdot10^{-3}$ \cite{Gra.83}. Comparison with the yield
of $f_2$ production
\cite{May.90} in the $\bar{p}p\to \pi^0 f_2$ channel of
$(3.9\pm1.1)\cdot10^{-3}$
shows that the present experimental information on the ratio
$f'_2/f_2$ ratio is rather inconclusive.
\vskip0.3cm

 Since the $f'_2$ is a spin--triplet
P--wave state in the na\"{\i}ve quark model, the type of argument we used
to motivate enhancement of \f\ production in $^3S_1$ \p\ \an s would
favour a large
$f'_2/f_2$ ratio in $^3P_1$ \p\ \an s.
It is interesting to note that the $f_2$ yield in P--wave
\p\ \an\ is known to be
five times greater than in the S-wave:
$Y_{f_2}^{P}=1.85\pm0.24 \%$ \cite{May.90}.
If the above prediction of enhanced
 $f'_2$ production is correct, and the effect is as large as in
the $^3S_1$ \f\ production case,
the signal for $f'_2$ production in P--wave \p\ \an\  should be clearly
visible,
with the branching ratio of $\bar{p}p\to \pi^{0} f'_2$ possibly
as large as 0.1-0.2 \%.
    This suggestion  could be tested in data  recently taken  by the
OBELIX collaboration on \p\ annihilation in gaseous
hydrogen at small pressure.
    There are already some data on  $f_2'$ production in \p\
annihilation in flight \cite{Che.77},\cite{Vui.76} indicating
substantial apparent violation of the na\"{\i}ve OZI prediction
(\ref{f1})-(\ref{f2}), but the statistics in these experiments was
scarce, and  new high--statistics data are needed.
\vskip0.3cm

        Ref. \cite{Lev.94} considered
production of $f'_2$ in the $\bar{p}p \to f'_2 \pi^0$ reaction
via final-state $K^*K$ and $\rho\pi$ interactions.
The calculated production rates of $f'_2$ from the S or
P states were rather small, of the order $10^{-6}$. This means that, if
any deviation from the na\"{\i}ve OZI rule is established for $f'_2$, it
could not be explained by rescattering.

\vspace {0.3cm}

2.\ Since annihilation at higher energies involves an increasing mixture
of partial waves, the S-wave state ``rearrangement" that favours,
for example, $\phi\pi$ state production from the $^3S_1$ initial state,
becomes more diluted (see the discussion in Sect.5). We expect therefore
that the $\phi\pi/\omega\pi$ ratio measured in annihilation in flight
will decline, following the decreasing
admixture of the $^3S_1$ state. Recent preliminary results from
the Crystal Barrel experiment \cite{Wied94} indicate that the branching
ratio of the
$\phi\pi^0$ channel decreases approximately 5 times when the momentum
of the antiproton is increased to $600\ MeV/c$ whereas the branching
ratio of the $K^*K$ stays constant. This is consistent with the fact that
 at $600\ MeV/c$ the percentage
of S-wave in $p\bar{p}$ annihilation is about $14-20\%$ \cite{Mar.92}.
\vskip0.3cm

3.\ It is important to check the spin dependence of the OZI-violating
$\bar{p}p\to \phi\phi$ amplitude. Our model predicts that an even larger
deviation from the na\"{\i}ve OZI rule should be seen in an experiment with
polarized beam and target
when an initial spin-triplet state is prepared. In a spin-singlet state
the apparent OZI violation should, according to our model, be less pronounced.
\vskip0.3cm

4.\ An interesting possibility for testing the model is provided by the
$\phi\pi\pi$ final state where, contrary to two-particle channels for $\phi$
production, annihilation in the same partial wave is possible from both
the spin-triplet and spin-singlet states. A spin-parity analysis of the
Dalitz plot for $\bar{p}p\to \phi\pi\pi$ annihilation should demonstrate
dominance of the $^3S_1$ initial state, if our hypothesis is correct.
\vskip0.3cm

5.\ It is interesting to study the spin structure of the OZI-allowed
process
\begin{equation}
\bar{p} + p \to K^* + \bar{K}^*
\end{equation}
If intrinsic strangeness manifests itself also in OZI-allowed
processes, spin correlations should appear in the final state.
For example, when the initial $\bar{p}p$ pair is in the spin-triplet state,
the final $K^* \bar{K}^*$ channel should be dominated by the $S=2$ state.
\vskip0.3cm

6.\ The largest violation of the na\"{\i}ve OZI rule occurs in the $\phi\gamma$
channel (see Table 2). This channel was measured for antiproton
annihilation in liquid, where S-wave annihilation is dominant.
The $\phi\gamma$ final state is possible either from spin-singlet
$^1S_0$ or from spin-triplet $^3P_{0,1,2}$ states. If
the $\phi$ production is really enhanced for spin-triplet states,
then one would expect that the ratio $\phi\gamma/\omega\gamma$
will increase for annihilation in hydrogen gas at normal temperature and
pressure or at low pressure, where the P-wave annihilation is dominant.

\vskip0.3cm

7.\  Another possible test of the \f\ production mechanism is
 provided by its final--state decay  angular distribution, as measured in the
reaction
\begin{eqnarray}
p + p &\longrightarrow &\phi + X,\\
& &\hookrightarrow e^{+}e^{-} \label{dec}
\end{eqnarray}
for example. Since our production mechanism proceeds via a vertex
$ (\bar{s}\gamma_{\mu}s) \phi_{\mu}$,
where the $s$ and $\bar{s}$  come from the initial--state proton
wavefunctions, we expect an angular distribution
\begin{equation}
W(\theta) = 1 + cos^2\theta
\end{equation}
for the $e^+e^-$ pair,
where $\theta$ is the angle of the $e^-$ ($e^{+}$) relative to the incident
proton beam direction measured in the rest frame of the $\phi$.
\vspace {0.3cm}

8.\ An interesting test of the \f\ production mechanism involves
an application of the constituent counting rules for high momentum--transfer
collisions \cite{Mat73},\cite{Bro73}:
\begin{equation}
{{d\sigma} \over {dt}}_{\theta_{cm}\ fixed}^{(A+B \to C+D)} =
{1 \over {s^{n_A + n_B + n_C + n_D-2}}}\ f(s/t),  \label{qcr}
\end{equation}
where $n_{A,B,C,D}$ are the numbers of constituents participating in the
hard collision. Consider for definiteness the processes $\pi p \to \phi n$
and $\bar{p}p \to \phi\pi$, compared with $\pi p \to \omega n$
and $\bar{p}p \to \omega \pi$. The latter reactions can proceed via
valence quarks in the participating hadrons, in particular, via the \=qq
component of the $\omega$, and the differential cross--sections should
therefore fall off  as $s^{-8}$. On the other hand, if
$\phi$ mesons are produced entirely via the $|uud s\bar{s}>$
components of the nucleon wave functions, the \f\ differential
cross--sections should fall off as
$s^{-12}$, resulting in a disappearance of the apparent OZI violation
at high energies. On the other hand, if  $\phi$ production
proceeds via a three-gluon intermediate state,
this can result in an  $s^{-9}$ behaviour of the differential
cross--section.
To be conclusive, such a  quark counting rule analysis should only be made
at energies where the validity of (\ref{qcr}) is  established,
e.g. above 10 GeV.
However, qualitative features of this type may also appear at low
energies.
The trend for the na\"{\i}ve OZI rule (\ref{R}) to become more accurate
at high energies (see Fig.2) and the different $t-$dependences in
$\pi p\to\phi n$ and $\pi p\to\omega n$ shown in Fig.5 are clearly
consistent with the
expectations of the quark counting rules, even though the latter are
not strictly applicable.
\vskip0.3cm

9.\ The studies of
angular distributions of \f\ and $\omega$ production
for annihilation in flight $\bar{p}p \to \phi(\omega) \pi$
are interesting since these distributions
should
be different if \f\ and $\omega$ production mechanisms are not the same.
Such a feature
was already noted \cite{Coo.78}
in a bubble chamber experiment on $\bar{p}p \to \phi \pi\pi$
and $\omega\pi\pi$, albeit with low statistics.
\vskip0.3cm

10.\ The largest momentum transfer in $\phi$ production by stopped
antiproton annihilation is available in the so-called Pontecorvo reaction
\begin{equation}
\bar{p} + d \to \phi + n
\end{equation}
We may therefore expect very high $\phi/\omega$ ratios in reactions
of this type.

\vskip2.5cm

{\bf Acknowledgments}

    We thank M.Alberg, C.Amsler, A.Grigoryan, E.Klempt, R.Landua, F.Lehar,
C.Strass\-burger and U.Wiedner for useful remarks.

The research of M.K. was supported in part by grant No. 90-00342 from the
US-Israel Science Foundation and by the Basic Research Foundation
administered by the Israel Academy of Sciences.
   The work of D.E.K. was supported in part by Bundesministerium f\"{u}r
Forschung und Technologie under grant 06-BI-721 and by INTAS Association.
    M.G.S. acknowledges the support from the Russian
Fund of Fundamental Research under grant No. 93-02-3997 as well as
from the International Science Foundation, grant ML9000.

\newpage
Table 1. The ratios $R=\phi X/\omega X $ for production
of the $\phi$ and $\omega$ - mesons in $pp$, $\bar p p$ and $\pi p$
interactions at $P_L$ different from zero. The parameter Z of the OZI-rule
 violation is
calculated for $\delta=\Theta-\Theta_i=3.7^0$, assuming
identical phases of the $\phi$ and $\omega$ production amplitudes.  \\~\\
\begin{tabular}{crclll}
\hline
Initial state & $P_L$ (GeV/c)&Final state X
& $R=\phi X/\omega X \cdot 10^{3}$
&$\left| Z\right|~(\%)$ & Refs.\\ \hline
$\pi^+ n$ & 1.54-2.6 & p & $21.0\pm11.0$
& $8\pm4$ &\cite{Dav.70},\cite{Dan.70} \\
$\pi^+ p$ & 3.54     & $\pi^+p$ & $19.0\pm11.0$
& $7\pm4$ & \cite{Abo.63} \\
$\pi^- p$ & 5-6      & n        & $3.5\pm1.0$
&  $0.5\pm0.8$ &\cite{Ayr.74} \\
$\pi^- p$ & 6        & n        & $3.2\pm0.4$
&  $0.8\pm0.4$ & \cite{Coh.77}\\
$\pi^- p$ & 10       & $\pi^-p$  & $6.0\pm3.0$
&  $1.3\pm2.0$ & \cite{Bal.77}\\
$\pi^- p$ & 19       & $2\pi^-\pi^+p$ & $5.0^{+5}_{-2}$
&  $0.6\pm2.5$&\cite{Woo.76} \\
$\pi^- p$ & 360     & X          & $14.0\pm6.0$
&  $5\pm3$ & \cite{Agu.89}  \\
\hline
pp        & 10       & pp & $20.0\pm5.0$
&  $8\pm2$ & \cite{Bal.77}\\
pp        & 24       & pp & $26.5\pm18.8$
&  $10\pm6$ & \cite{Blo.75}  \\
pp        & 24       & $\pi^+\pi^-pp$ & $1.2\pm0.8$
&  $3\pm1$ & \cite{Blo.75} \\
pp        & 24       & pp m$\pi^+\pi^-$, m=0,1,2
& $19.0\pm7.0$ & $7\pm3$ & \cite{Blo.75}\\
pp        & 360     & X          & $4.0\pm5.0$
&  $0.1\pm4$ & \cite{Agu.91}  \\
\hline
$\bar p p$ & 0.7      &$\pi^+\pi^-$ & $19.0\pm5^{*)}$
&  $7\pm2$ &\cite{Coo.78} \\
$\bar p p$ & 0.7      & $\rho^0$      & $13.0\pm4^{*)}$
& $5\pm2$ &\cite{Coo.78} \\
$\bar p p$ & 1.2      &$\pi^+\pi^-$ & $11.0\pm^{+3}_{-4}$
&  $4\pm1$ &\cite{Don.77}  \\
$\bar p p$ & 2.3      &$\pi^+\pi^-$ & $17.5\pm3.4$
&  $7\pm1$ &\cite{Che.77}\\
$\bar p p$ & 3.6      & $\pi^+\pi^-$ & $9.0^{+4}_{-7}$
&  $3\pm3$ &\cite{Don.77}\\
\hline\\~\\
\end{tabular}

$^{*)}$ corrected for phase space.\\~\\
\newpage
TABLE 2. The ratios $R=\phi X/\omega X $ for production of
the $\phi$ and $\omega$ - mesons in
antinucleon annihilation at rest.
The parameter Z of the OZI-rule violation is
calculated as in Table 1. The data are given for annihilation in
liquid hydrogen target (percentage of annihilation from P-wave
is $\sim 10-20 \%$),
gas target ($\sim$61\% P-wave) and
 LX-trigger \cite{Rei.91} ($\sim$86-91\% P-wave).
 \\~\\
\begin{tabular}{llllll}
\hline
Final state& Initial states & B.R.$\cdot10^{4}$& $R\cdot 10^{3}$
&$\left| Z\right|~(\%)$ &Comments\\
\hline
$\phi\gamma$ & $^1S_0,^3P_J$ & $0.17\pm0.04$  & $250\pm89$
&  $42\pm8$ & liquid,\cite{Fae.93}\\
\hline
$\phi\pi^0$ & $^3S_1,^1P_1$ &$5.5\pm0.7$  & $96\pm15$
&  $24\pm2$ & liquid,\cite{Fae.93}\\
$\phi\pi^0$ & &$1.9\pm0.5$ &
&   & gas, \cite{Rei.91}\\
$\phi\pi^0$ & &$0.3\pm0.3$ &
&   & LX-trigger, \cite{Rei.91}\\
\hline
$\phi\pi^-$ &$^3S_1,^1P_1$ &$9.0\pm1.1$
& $83\pm25$
&  $22\pm4$ & liquid,\cite{Biz.74}-\cite{Bet.67}\\
$\phi\pi^-$ & &$14.8\pm1.1$  & $133\pm26$
&  $29\pm3$ & $\bar p d,
p<200~ MeV/c$, \cite{Abl.93} \\
$\phi\pi^-$ & & & $113\pm30$
&  $27\pm4$ & $\bar p d,p>400~ MeV/c$, \cite{Abl.93} \\
$\phi\pi^+$ &  & & $110\pm15$
&  $26\pm2$ & $\bar n p$, \cite{Abl.93} \\
\hline
$\phi\eta$ &$^3S_1,^1P_1$ & $0.9\pm0.3$ & $6.0\pm2.0$
&  $1.3\pm1.2$ & liquid,\cite{Fae.93}\\
$\phi\eta$ & &$0.37\pm0.09$ &
&  & gas, \cite{Rei.91}\\
$\phi\eta$ & &$0.41\pm0.16$ &
&   & LX-trigger, \cite{Rei.91}\\
\hline
$\phi\rho$ & $^1S_0,^3P_J$ & $3.4\pm0.8$ & $6.3\pm1.6$
&  $1.4\pm1.0$ & gas, \cite{Rei.91},\cite{Wei.93}\\
$\phi\rho$ & &$4.4\pm1.2$ & $7.5\pm2.4$
&  $2.1\pm1.2$ & LX-trigger, \cite{Rei.91},\cite{Wei.93}\\
\hline
$\phi\omega$ & $^1S_0,^3P_{0,2}$ &$6.3\pm2.3$  & $19\pm7$
& $7\pm4$ & liquid, \cite{Biz.71},\cite{Ams.93}\\
$\phi\omega$ & &$3.0\pm1.1$  &
&          & gas, \cite{Rei.91}\\
$\phi\omega$ & &$4.2\pm1.4$  &
&          & LX-trigger, \cite{Rei.91}\\
\hline
$\phi\pi^0\pi^0$ &$^{1,3}S_{0,1},^{1,3}P_J$ &$1.2\pm0.6$  & $6.0\pm3.0$
& $1.3\pm2.0$ & liquid,\cite{Fae.93}\\
$\phi\pi^-\pi^+$ & &$4.6\pm0.9$  & $7.0\pm1.4$
& $1.9\pm0.8$ & liquid,\cite{Biz.69}\\
$\phi X,
X=\pi^+\pi^-, \rho$ & &$5.4\pm1.0$  &$7.9\pm1.7$
&$2.4\pm1.0$  & gas, \cite{Rei.91},\cite{Wei.93}\\
$\phi X,
X=\pi^+\pi^-, \rho$ & &$7.7\pm1.7$  &$11.0\pm3.0$
&$4.0\pm1.4$  & LX-trigger, \cite{Rei.91},\cite{Wei.93}\\
\hline\\~\\
\end{tabular}
\newpage
\medskip
{\bf FIGURE CAPTIONS}\\

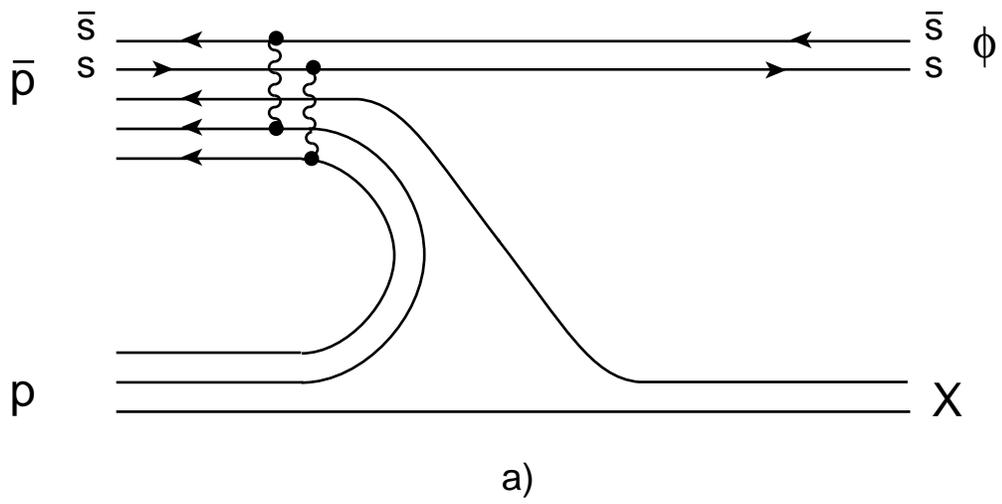
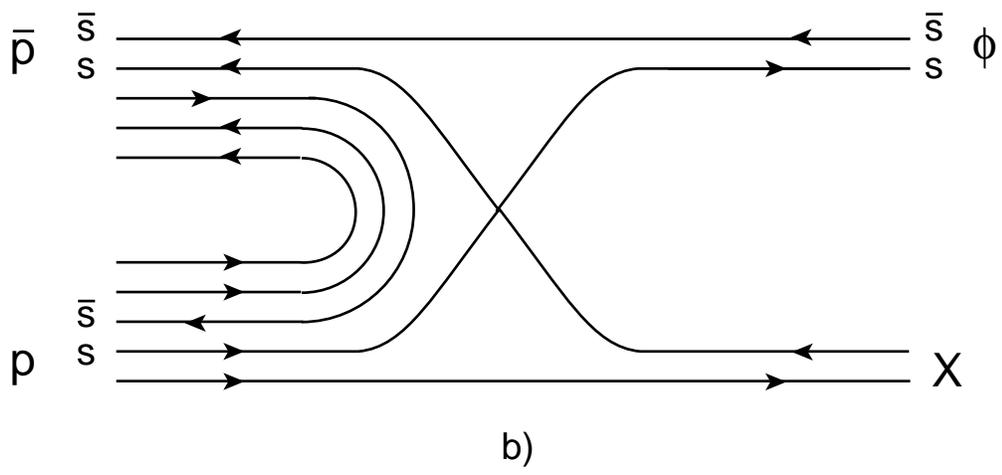
{\bf Figure 1.} a) Production of a $\phi$ meson in $\bar{p}p$  \an\ by the
OZI-allowed process of \=ss rearrangement from $|uud\bar{s}s>$
components of the proton wave function.
\vskip0.3cm
b) Shake-out of a $\phi$ meson from a $|uud\bar{s}s>$
component of the proton wave function.
\vspace {0.3cm}

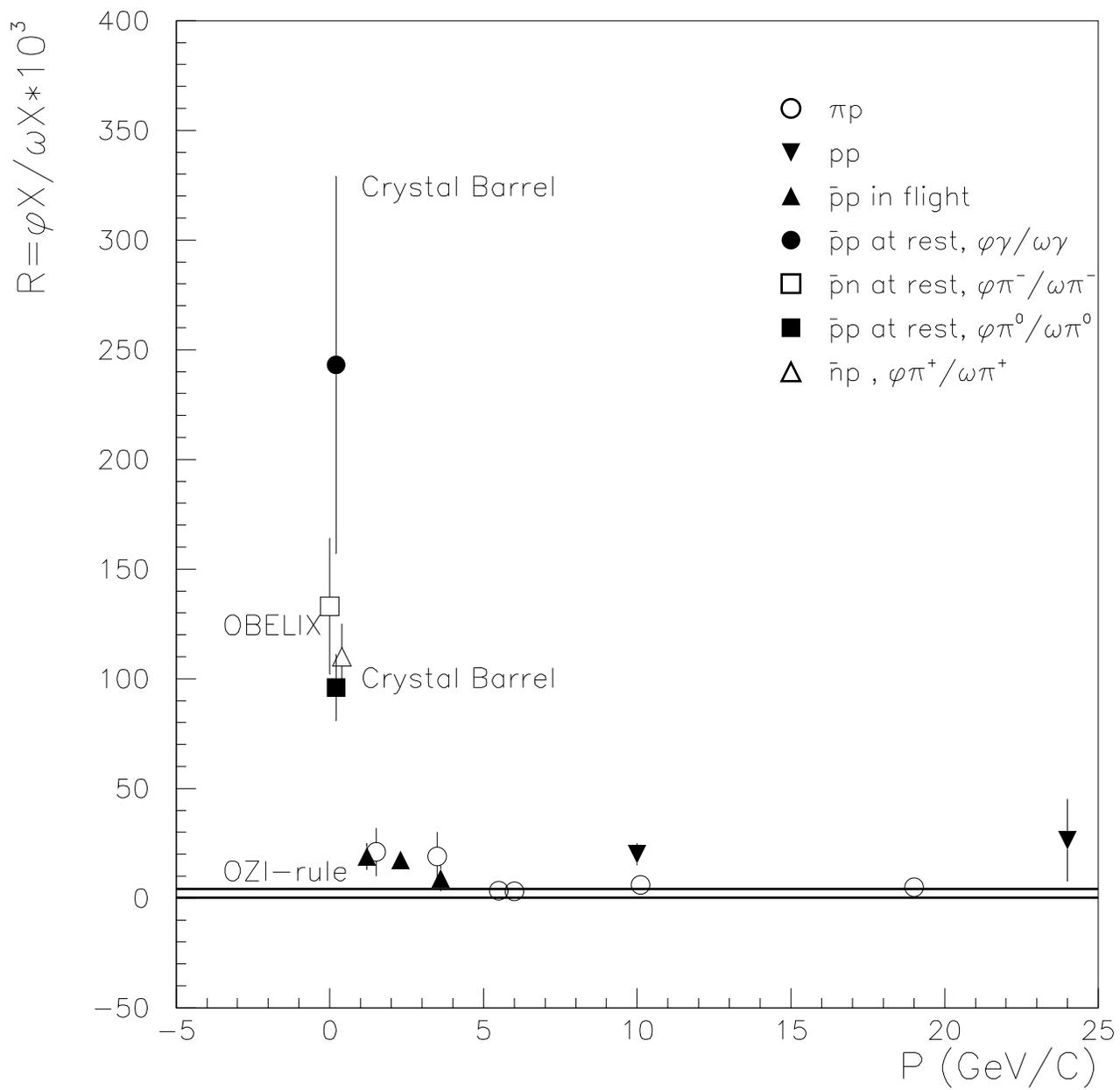
{\bf Figure 2.} Ratios $R=\phi X/\omega X$ in different reactions at
increasing momenta p.
\vspace {0.3cm}

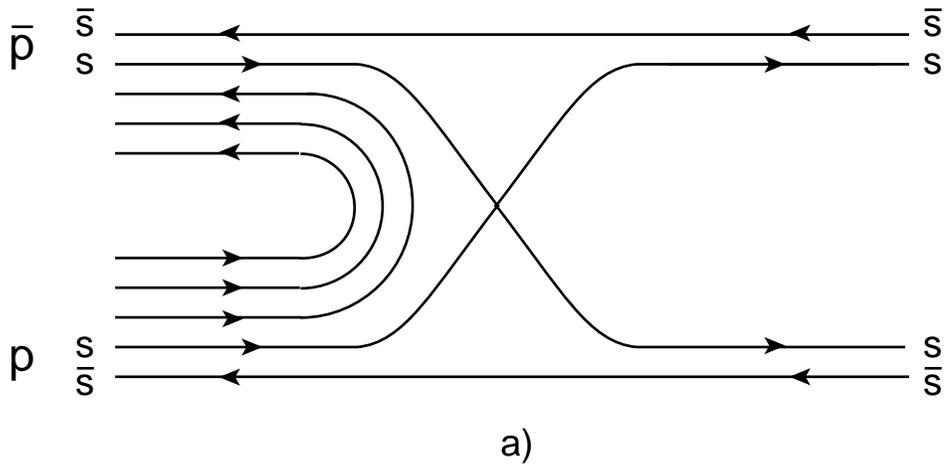
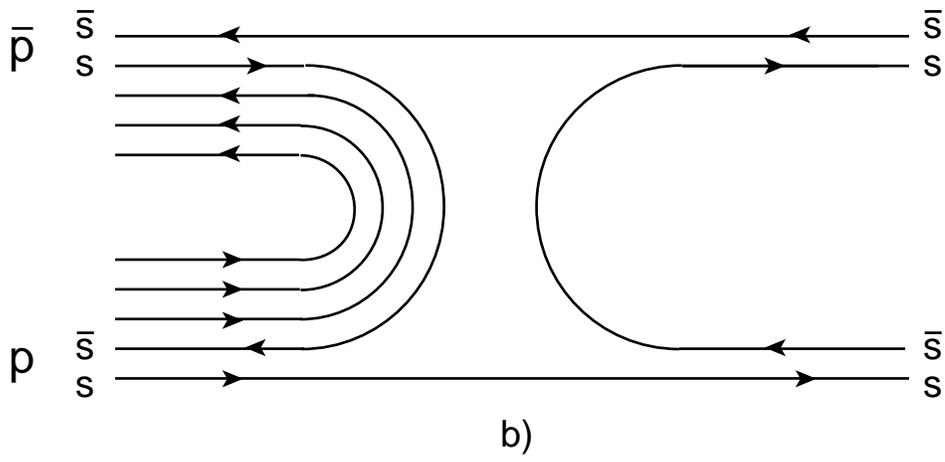
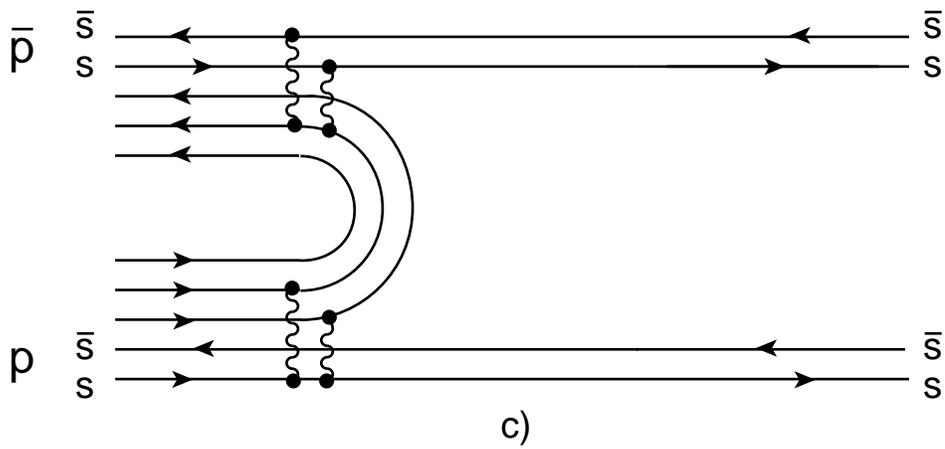
{\bf Figure 3.} Diagrams contributing
to the double production of \=ss mesons via a) double annihilation,
b) via creation of an additional \=ss pair, and c) via double
shake-out of \=ss pairs in $|uud\bar{s}s>$ components.

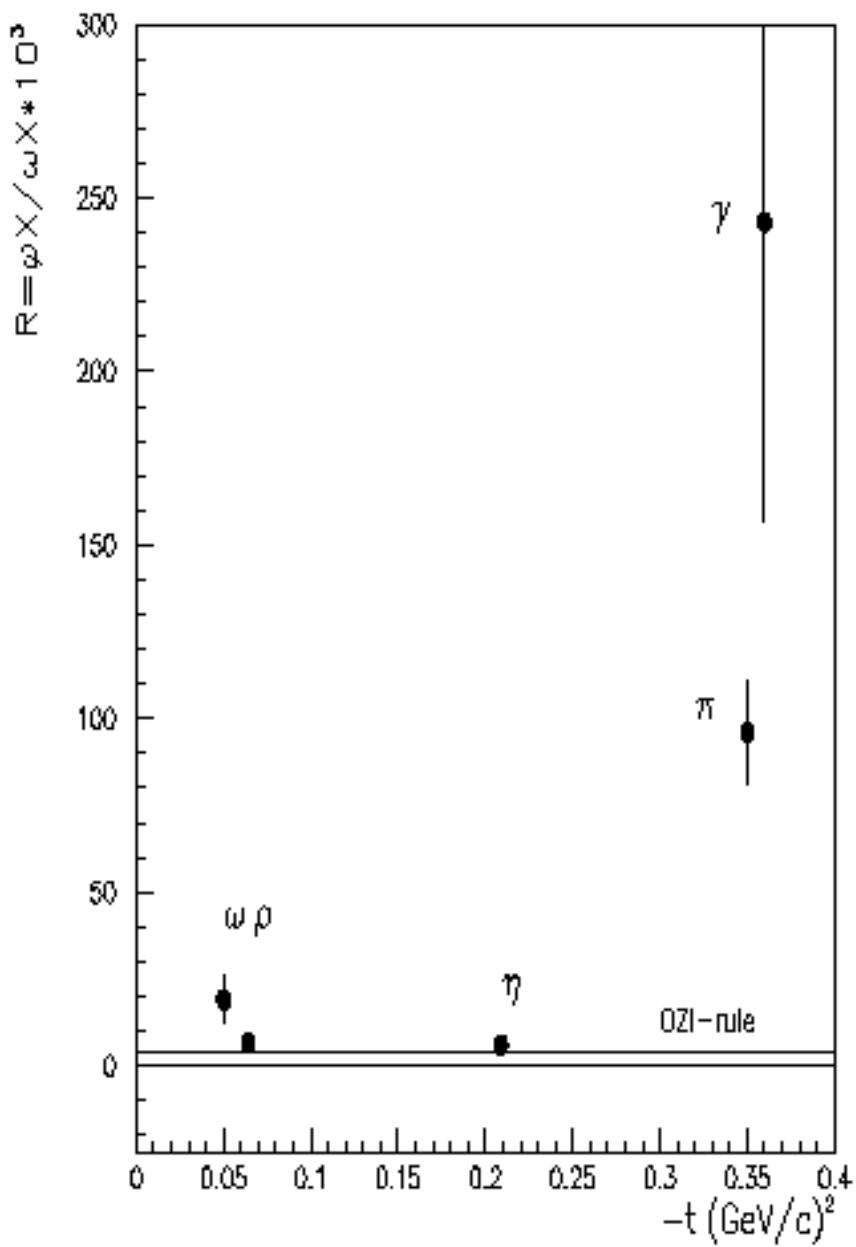
{\bf Figure 4.}  Dependence
of the ratios $R=\phi X/\omega X$ in different reactions
of stopped antiproton annihilation in hydrogen
on momentum transfer in $\bar{p}p\to \phi X$.
Experimental data are from Table 2.

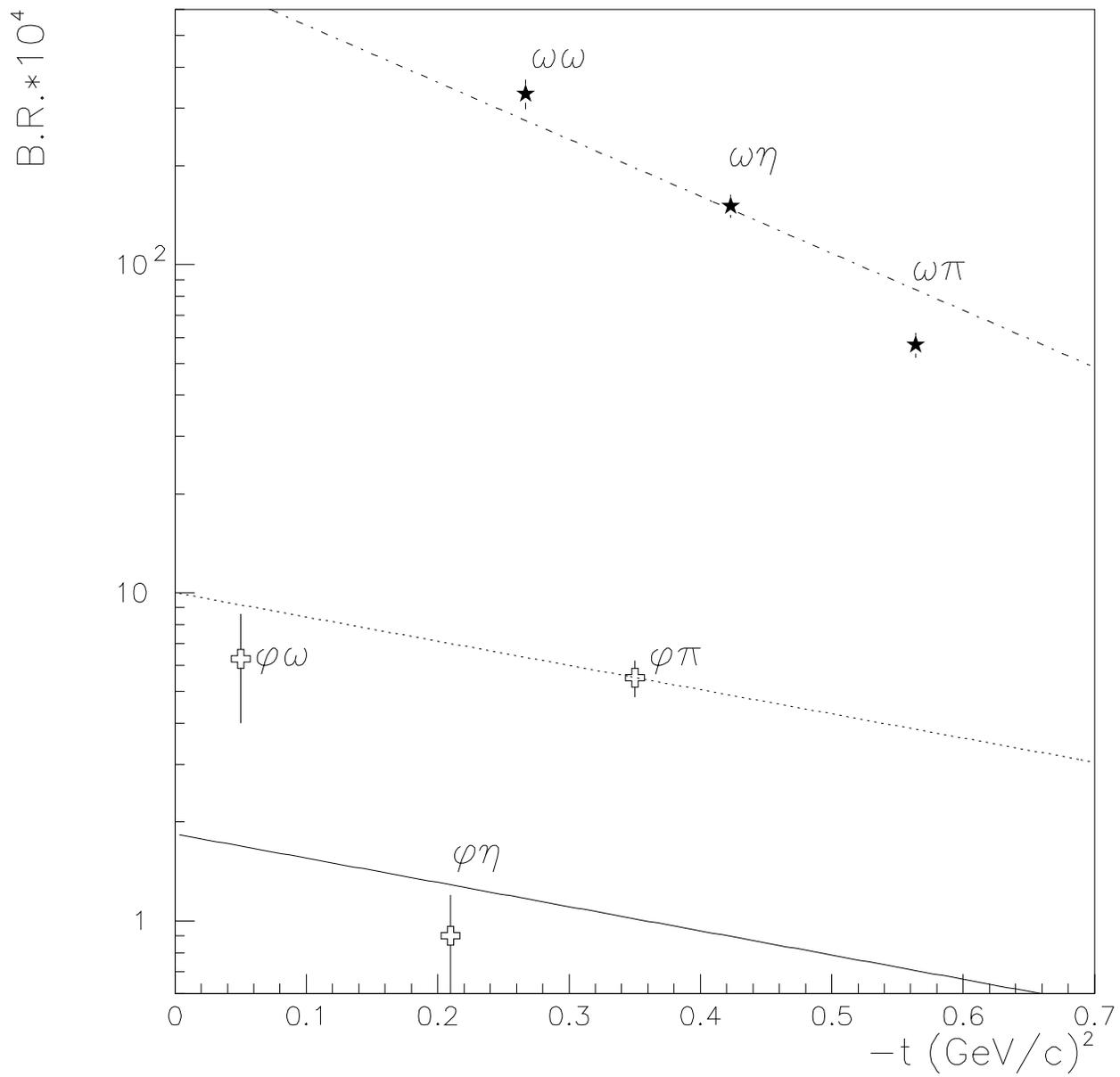
{\bf Figure 5.}  Dependences
of the branching ratios of $\phi X$ and $\omega X$ in different channels
of stopped antiproton annihilation in hydrogen
on momentum transfer.
Experimental data are from Table 2. The solid (lowest) and dash-dotted
(highest) lines are
the $t-$dependences measured in $\pi p\to \phi n$ and $\pi p\to \omega n$
reactions, respectively, with relative normalization fixed to be the same as
for the experimental cross
sections \cite{Coh.77}. The dotted line has the same functional dependence
as the solid line,
but is normalized to fit the $\bar{p}p\to \phi\omega,\phi\pi$ data.

\newpage
 \begin{thebibliography}{999}
\bibitem{Ell.89} J. Ellis, E. Gabathuler and M. Karliner,
Phys.Lett.{\underline{B217}} (1989) 173.
\bibitem{Ell.93}
J. Ellis and M. Karliner, Phys.Lett.{\underline{B313}} (1993) 131.
\bibitem{Dec.92}
R. Decker, M. Nowakowski and U. Wiedner,
Fort.Phys.{\underline{41}} (1993) 87.
\bibitem{2dim}
Y. Frishman and M. Karliner, Nucl.Phys.{\underline{B344}} (1990) 393;\\
J. Ellis, Y. Frishman, A. Hanany and M. Karliner,
Nucl.Phys.{\underline{B382}} (1992) 189.
\bibitem{GL.91}
J. Gasser, H. Leutwyler and M.E. Sainio, Phys.Lett.{\underline{B253}} (1991)
252, 260.
\bibitem{lat}
S.-J. Dong and K.-F. Liu, hep-lat/9412059, 1994.
\bibitem{OZI}
S. Okubo, Phys.Lett.{\underline{B5}} (1963) 165.\\
G. Zweig, CERN Report No.8419/TH412 (1964).\\
I. Iizuka, Prog. Theor. Phys. Suppl. 37 {\underline{38}} (1966) 21.\\
see also G. Alexander, H.J. Lipkin and P. Scheck, Phys.Rev.Lett.
{\underline{17}} (1966) 412.
\bibitem{Lip.76}
H.J.Lipkin, Phys.Lett.{\underline{B60}} (1976) 371.
\bibitem{Dav.70}
D.W. Davies et al., Phys.Rev.{\underline{D2}} (1970) 506.
\bibitem{Dan.70}
J.S. Danburg et al., Phys.Rev.{\underline{D2}} (1970) 2564.
\bibitem{Abo.63}
M. Abolins et al., Phys.Rev.Lett.{\underline{11}} (1963) 381.
\bibitem{Ayr.74}
D. Ayres et al., Phys.Rev.Lett.{\underline{32}} (1974) 1463.
\bibitem{Coh.77}
D. Cohen et al., Phys.Rev.Lett.{\underline{38}} (1977) 269.
\bibitem{Bal.77}
R. Baldi et al., Phys.Lett.{\underline{B68}} (1977) 381.
\bibitem{Woo.76}
P.L. Woodworth et al., Phys.Lett.{\underline{B65}} (1976) 89.
\bibitem{Agu.89}
The LEBC-EHS Collaboration, M. Aguilar-Benitez et al., Z.Phys.{\underline{C44}}
 (1989) 531.
\bibitem{Blo.75}
V. Blobel et al., Phys.Lett.{\underline{B59}} (1975) 88.
\bibitem{Agu.91}
The LEBC-EHS Collaboration,
M. Aguilar-Benitez et al., Z.Phys.{\underline{C50}} (1991) 405.
\bibitem{Coo.78}
A.M. Cooper  et  al.,  Nucl.Phys.{\underline{B146}} (1978) 1.
\bibitem {Don.77}
R.A. Donald et al., Phys.Lett.{\underline{B61}} (1976) 210.
\bibitem{Che.77}
C.K. Chen et al., Nucl.Phys.{\underline{B130}} (1977) 269.
\bibitem{Rei.91}
The ASTERIX Collaboration, J. Reifenrother et al., Phys.Lett.{\underline{B267}}
 (1991) 299.
\bibitem{Fae.93}
The Crystal Barrel Collaboration, M.A. Faessler et al.,
Proc. NAN-93 Conference, Moscow, 1993; Phys. At. Nuclei {\underline{57}} (1994)
1693.
\bibitem{Abl.93}
The OBELIX Collaboration, V.G. Ableev et al., Proc. NAN-93 Conference,
Moscow, 1993; Phys. At. Nuclei {\underline{57}} (1994)
1716.
\bibitem{Luc.93}
The OBELIX Collaboration, V.G. Ableev et. al.,
 Phys.Let., {\underline{B334}} (1994) 237.
\bibitem{Biz.74}
R. Bizzarri et al., Nuov.Cim.{\underline{A20}} (1974) 393.
\bibitem{Bet.69}
A. Bettini et al., Nuov.Cim.{\underline{A63}} (1969) 1199.
\bibitem{Biz.70}
R. Bizzarri et al., Phys.Rev.Lett.{\underline{25}} (1970) 1385.
\bibitem{Bet.67}
A. Bettini et al., Nuov.Cim.{\underline{A47}} (1967) 642.
\bibitem{Biz.69}
R. Bizzarri et al., Nucl.Phys.{\underline{B14}} (1969) 169.
\bibitem{Biz.71}
R. Bizzarri et al., Nucl.Phys.{\underline{B27}} (1971) 140.
\bibitem{Wei.93}
The ASTERIX Collaboration, P. Weidenauer et al., Z.Phys.{\underline{C59}}
(1993)
 387.
\bibitem{Ams.93}
The Crystal Barrel Collaboration, C. Amsler et al., Z.Phys.{\underline{C58}}
 (1993) 175.
\bibitem{EMC}
The EMC Collaboration, J. Ashman et al., Phys.Lett.{\underline{B206}}
(1988) 364.\\
The EMC Collaboration, J. Ashman et al., Nucl.Phys.{\underline{B328}}
(1989) 1.\\
The NMC Collaboration, P. Amaudruz et al., Phys.Lett.{\underline{B295}}
(1992) 159.\\
The SMC Collaboration, B. Adeva et al., Phys.Lett.{\underline{B302}}
(1993) 533.\\
The E142 Collaboration, P.L. Anthony et al., Phys.Rev.Lett.{\underline{71}}
(1993) 959.\\
The E143 Collaboration, K. Abe et al., SLAC-PUB-6508.
\bibitem{Bat.90}
The PS 179 Collaboration, Yu.A. Batusov et al., JINR preprint, E1-90-118, 1990,
 Dubna
\bibitem{Bal.66}
C. Baltay et al., Phys.Rev.{\underline{145}} (1966) 1103.
\bibitem{Tan.85}
T. Tanimori et al., Phys.Rev.Lett.{\underline{55}} (1985) 1835.
\bibitem{Has.92}
A. Hasan et al., Nucl.Phys.{\underline{B378}} (1992) 3.
\bibitem{Tan.90}
T. Tanimori et al., Phys.Rev.{\underline{D41}} (1990) 744.
\bibitem{Mac.93}
The JETSET Collaboration, M. Macri et al., Nucl.Phys.{\underline{A558}} (1993)
 27c.
\bibitem{Hoh76}
G. H\"ohler et al., Nucl.Phys.{\underline{B114}} (1976) 505.
\bibitem{Dub}
S. Dubnicka, Nuovo Cimento {\underline{A100}} (1988) 1.
\bibitem{Jaf89}
R.L. Jaffe, Phys.Lett.{\underline{B229}} (1989) 275.
\bibitem{Ast.88}
D. Aston et al., Phys.Lett.{\underline{B215}} (1988) 799.
\bibitem{Dov89}
C.B. Dover and P.M. Fishbane, Phys.Rev.Lett.{\underline{62}} (1989) 2917.
\bibitem{Bit.87}
The Lepton-F Collaboration, S.I. Bityukov S.I. et al.,
Phys.Lett.{\underline{B188}} (1987) 383.
\bibitem{Lan.88}
L.G. Landsberg, Sov.J.Part.Nucl.{\underline{21}} (1990) 446;
Preprint IHEP, 88-143, Protvino, 1988.
\bibitem{Bra.92}
The Crystal Barrel Collaboration, K. Braune et al.,
Nucl.Phys.{\underline{A558}} (1993) 269c.
\bibitem{Loc.93}
M.P. Locher, Y. Lu and B-S. Zou
Z.Phys.{\underline{A347}} (1994) 281.
\bibitem{Buz.93}
D. Buzatu and F. Lev, Phys.Lett.
{\underline{B329}} (1994) 143.
\bibitem{Lip.91}
H.J. Lipkin, Int. J. Mod.Phys.{\underline{E1}} (1992) 603.
\bibitem{Gei.91}
P. Geiger and N. Isgur, Phys.Rev.Lett.{\underline{67}} (1991) 1066.
\bibitem{Kon.93}
K. K\"onigsmann, Preprint CERN-PPE/93-182, Geneva, 1993; Proc. PANIC 93,
Perugia, 1993, Ed. A. Pascolini; World Scientific, 1994.
\bibitem{Iof.90}
B.L. Ioffe and M. Karliner, Phys.Lett.{\underline{B247}} (1990) 387.
\bibitem{Hen.92}
E.M. Henley, G. Krein and A.G. Williams, Phys.Lett.{\underline{B281}}
(1992) 178.
\bibitem{Ell.93a}
J. Ellis, M.Karliner, Preprint CERN-TH.7324/94, Geneva, 1994, hep-ph/9407287,
Phys.Lett.B, to appear.
\bibitem{Dov.92}
C.B. Dover et al., Prog. Part. Nucl. Phys.{\underline{29}} (1992) 87.
\bibitem{May.90}
The ASTERIX Collaboration, B. May et al., Z.Phys.{\underline{C46}} (1990) 191;
 203.
\bibitem{Bar.93}
The PS185 Collaboration, P. Barnes et al., Nucl.Phys.{\underline{A558}} (1993)
 277c.
\bibitem{Gra.83}
L. Gray et al., Phys.Rev.{\underline{D27}} (1983) 307.
\bibitem{Vui.76}
V. Vuillemin et al., Nuov.Cim.{\underline{A33}} (1976) 133.
\bibitem{Lev.94}
D. Buzatu and F. Lev, JINR preprint, E4-94-158, Dubna, 1994.
\bibitem{Wied94}
The Crystal Barrel Collaboration, U. Wiedner et al., Proc. LEAP'94 Conf., Bled,
 1994, to appear.
\bibitem{Mar.92}
M.Maruyama, Proc. LEAP'90 Conf., Stockholm, 1990, p.3.
\bibitem{New.93}
Nouvelles de Saturne, {\underline{17}} (1993) 59.
\bibitem{Mat73}
V.A. Matveev, R.M. Muradyan and A.N. Tavkhelidze, Nuovo Cim. Lett.
{\underline{7}} (1973) 719.
\bibitem{Bro73}
S.J. Brodsky and G.R. Farrar, Phys.Rev.Lett.{\underline{31}} (1973) 1153.
\end{thebibliography}
\end{document}